# Tunnel Electron Radiation on Recombination Center

P.A. Golovinski [1,2]*, A.A. Drobyshev [2]

[1] *Moscow Institute of Physics and Technology (State University), Moscow, Russia*
[2] *Voronezh State University of Architecture and Civil Engineering, 394006, Voronezh, Russia*

The photorecombination radiation on a neighbor center of an electron ionized in the nonlinear tunneling regime is discussed. In the framework of the active electron model an analytical solution of the problem has been obtained. The analytical solution for distribution of the   radiation energy spectrum has been derived both for one and three dimension problems. The optimal parameters of the field and the distance between quantum wells for a process of effective photorecombination with a broad spectrum have been formulated.

*Introduction.* – During the past decade the vast efforts have been concentrated on the dynamical electron effects of the high harmonic generation in the focus of the femtosecond and attosecond laser pulses [1-3]. The nonlinear ionization of atoms with the following acceleration of the electrons by a monochromatic low frequency laser field with high intensity can result in the recombination of an electron in the nearest capture centers and is accompanied by radiation of the short time electromagnetic pulses. In this work we discuss radiation of an electron, tunneled from the bound quantum state under the action of a low frequency electric component of the laser field $F(t) = F_0 \cos(\omega t)$ ( $F_0$ is the amplitude of an electric strength, $\omega$ is the laser field frequency),    in the process of the next center recombination.

The low frequency limit for a nonlinear ionization from the state with the energy  $E_i$ is realized when the Keldysh's parameter $\gamma = \omega \sqrt{2|E_i|} / F_0 << 1$ [4]. In this limiting case the ionization is a quasisteady process, and it mainly occur near the amplitude value $F_0$ of the laser field [5]. The validity of the model is confirmed by applicability of the ADK formula [6] to the description of nonlinear atom ionization. We describe electron radiation as a three step process: nonlinear ionization, acceleration in the laser field, and photorecombination to the final bound state [7, 8].

---

* golovinski@bk.ru

The majority of modern experiments for the high harmonic generation are in a good agreement with the Corcum's rescattering model [9, 10]. According to this model the radiation is a result of the electron, driving by alternating laser field, is returning to the initial atom. In some cases an electron comes back to the parent atom with the enough high energy and produces photons during the recombination. The probability of the process at a low frequency limit tends to zero as a result of wave packet dispersion when the time of a comeback is too large for efficient interaction with the parent center [11].

However, another process accompanied by photorecombination is possible and may be more efficiently. The laser field can be assumed to be quasi static, if the oscillation amplitude $F_0 / \omega^2$ be greater than the distance $R$ between two centers. We use atomic units $e = m = \hbar = 1$ throughout the paper. The collision energy of the electron then is estimated as $p^2 / 2 = R F_0$ . The recombination probability tends to maximum value $p \sim 1/a$, where $a$ is a characteristic size of the final state wave function. Hence, the maximum of the radiation intensity is achieving at the distance $R \sim (2 F a^2)^{-1}$ , and for $a \sim 1$, $I = 10^{12}$ W/cm$^2$ the corresponding optimal distance $R = 50$ a.u..

*Active electron model.* – We consider the active electron model based on the Feinman's propagation function [12]. The figure 1 demonstrates a scheme of photorecombination for the tunneling electron.

To describe the process we start from the Schrödinger's equation

$$\left( i \frac{\partial}{\partial t} - H \right) \psi(x,t) = W \psi(x,t), \tag{1}$$

where $H$ is the Hamiltonian of an electron in the external field, and

$$W(x,t) = W(x)e^{i\Omega t}, W(x) = i\left(\frac{2\pi\Omega}{V}\right)^{1/2}(\mathbf{e}_\alpha \mathbf{r}) \qquad (2)$$

is the operator of the one photon spontaneous emission in the dipole approximation [13]. Here $\Omega$ is the spontaneous photon frequency, $\mathbf{e}_\alpha$ is its polarization, $V$ is the quantization volume.

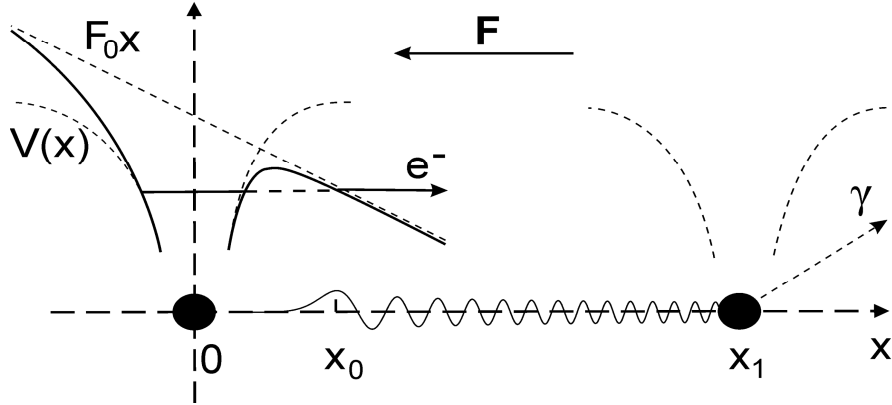

**Fig. 1.** The scheme of photorecombination for the tunneling electron.

The exact solution to the Schrödinger's Eq. (1) has to satisfy a linear homogeneous integral equation of the form

$$\psi(x,t) = i\int d^3x_1\, G(x,t;x_1,t_1)\psi_i(x_1,t_1), t > t_1, \qquad (3)$$

where $G(x,t;x_1,t_1)$ is the propagator or the Green's function, and the integration extends over the whole space. The probability amplitude for transition from the initial state $\psi_i(x,t)$ to the final state $\psi_f(x,t)$ has the form

$$S_{if} = i\int d^3x\, d^3x_1\, \psi_f^*(x,t)\, G(x,t;x_1,t_1)\psi_i(x_1,t_1). \qquad (4)$$

The Green's function for the time independent Hamiltonian (in the static fields) has eigenmod expansion

$$G_0(x,t;x_2,t_2) = -i\theta(t-t_2)\sum_\mu \exp(-iE_\mu(t-t_2))\varphi_\mu(x)\varphi_\mu^*(x_2), \qquad (5)$$

where $\varphi_n(x)$ are is the eigenfunctions of the operator $H$. Then, we have

$$S_{if} = \int d^3x_2\, dt_2\, d^3x_1\, \varphi_f^*(x_2)e^{i(E_f+\Omega)t_2}\, W(x_2)\, G_0(x_2,t_2;x_1,t_1)\psi_i(x_1,t_1). \qquad (6)$$

The analytical approach to the high harmonic generation in the delta potential model has been developed in [14], where the three step approximation has been substantiated. We operate with this approximation and represent the wave function of the active electron $i\int d^3x_1\, G_0(x_2,t_2;x_1,t_1)\psi_i(x_1,t_1)$ as a sum of the bound state $\exp(-iE_it_2)\varphi_i(x_2)$ and the wave packet $\psi_c(x_2,\,t)$ in the continuum [15], i.e.

$$\psi(x_2,\,t_2) = \exp(-iE_it_2)\varphi_i(x_2) + \psi_c(x_2,\,t_2). \qquad (7)$$

The propagation of this wave packet is described by the relation

$$\psi_c(x_2,\,t_2) = \sqrt{w(F_0)}\,G(x_2,t_2,x_0,0), \qquad (8)$$

where $G(x_2,t_2,x_0,0)$ is the Green's function of the electron in a homogeneous electric field $F_0$, $x_0$ is the turn point coordinate in the static potential at the moment $t = 0$. The expression for the $S$-matrix element we can rewrite in terms of the retarded Green's function for electric field

$$G(x_2,t_2,x_0,0) = -i\theta(t_2)\sum_{\mu}\exp(-iE_\mu t_2)\widetilde{\varphi}_\mu(x_2)\widetilde{\varphi}_\mu^*(x_0), \qquad (9)$$

where $\widetilde{\varphi}_\mu(x_2)$ is the electron wave function in a uniform electric field, as follows

$$S_{if} = \sqrt{w(F_0)}\int d^3x_2\, dt_2\, \varphi_f^*(x_2)e^{i(E_f+\Omega)t_2}\,W(x_2)G(x_2,t_2,x_0,0). \quad (10)$$

We, mostly, fully took into account the influence of the electric field on the continuum spectrum states. The summation in the Eq. (9) is replaced by the integration over continuum quantum numbers. After substitution of the Eq. (9) into the Eq. (10) we get

$$\left|S_{fi}\right|^2 = 2\pi w(F_0)\left|\sum_{\mu}\varphi_f(x)\big|W(x)\big|\widetilde{\varphi}_\mu(x)\widetilde{\varphi}_\mu^*(x_0)\right|^2\delta\big(E_f+\Omega-E_\mu\big)\cdot t. \qquad (11)$$

This equation contains the conservation energy law in the form of the delta function. The probability of transition per unit of time can be obtained by additional summation over final states. In this case it is a number of final states $dN$ for the radiated photons with the momentum in the interval $(p,\,p+dp)$, the polarization $\mathbf{e}_\alpha$ in some solid angle $do$, and the volume $V$:

$$dN = \frac{Vp^2 \, dp \, do}{c^2(2\pi)^3}. \tag{12}$$

The distribution density for the radiation energy over frequency we write down in the form

$$d\mathrm{E}_\Omega = w(F_0)\frac{4\Omega^4}{3c^3}|\mathbf{r}_\Omega|^2 \frac{d\Omega}{2\pi}. \tag{13}$$

This formula can be applied to the problems of an arbitrary dimension.

*Two quantum wells.* – As an application of the general formalism we consider a system with two parabolic quantum wells, which is the one dimensional problem. The sum in the Eq. (11) is changed to the integration over the energy, and

$$G(x,t,x_1,0) = \int\limits_{-\infty}^{\infty} dE \exp(-iEt)\varphi_E(x)\varphi_E(x_1), \tag{14}$$

where $\varphi_E(x)$ are normalized wave functions of the electron in a homogeneous electric field expressed in terms of the Airy function [18]:

$$\varphi_E(x) = \frac{\alpha}{\sqrt{F}}\mathrm{Ai}(-q),\, \mathrm{Ai}(-q) = \frac{1}{\pi}\int\limits_0^\infty \cos\left(\frac{u^3}{3} - uq\right) du, \tag{15}$$

$$q = \left(x + E/F_0\right)\alpha,\, \alpha = \left(2F_0\right)^{1/3}.$$

The Fourier components of the dipole matrix element are

$$D_\Omega = \sqrt{w(F_0)}\left\langle \varphi_f(x)|x|\varphi_{E_f+\Omega}(x)\right\rangle \varphi_{E_f+\Omega}(x_0). \tag{16}$$

The final state we take in the form of wave function for the quantum linear oscillator [19]

$$\varphi_f(x) = \left(\omega_0/\pi\right)^{1/2}\exp\left(-\omega_0(x - x_1)^2/2\right), \tag{17}$$

where $x_1$ is a recombination center of the electron.

The probability of ionization, per half of period of the external laser field oscillation $T/2 = \pi/\omega$, exponentially depends on the electric strength:

$$w(F_0) = \exp\left(-\frac{4\sqrt{2}|E_i|^{3/2}}{3F_0}\right)\frac{\pi}{\omega}. \tag{18}$$

The complete expression for the dipole matrix element results through Eq.(16) – Eq. (18):

$$D_\Omega = \frac{\pi \omega_0^{1/2} \alpha^2}{\omega F_0} \exp\left(-\frac{4\sqrt{2}|E_i|^{3/2}}{3F_0}\right)^{1/2} \mathrm{Ai}(-q_0) \times$$

$$\times \int_{-\infty}^{\infty} \mathrm{Ai}\left(-\left(x + E/F_0\right)\alpha\right) \cdot x \cdot \exp\left(-\omega_0(x - x_1)^2/2\right)dx,$$

$$q_0 = \left(x_0 + E/F_0\right)\alpha. \tag{19}$$

The coordinate $x_0 = -|E_i|/F_0$ corresponds to the return point. The Eq. (19) in combination with the Eq. (13) allows us to calculate the spectrum of the electron photorecombination. The main influence factor of modulation is the function $\left|\mathrm{Ai}\left(-(E - |E_i|)/(\alpha F_0)\right)\right|^2$. The integral in the Eq. (19) can be evaluated asymptotically for large arguments of the Airy function.

In the figure 2 two photorecombination spectrums are plotted for different intensities. The comparison of them shows decreasing of the intensity and increasing the number of oscillations in the spectrum.

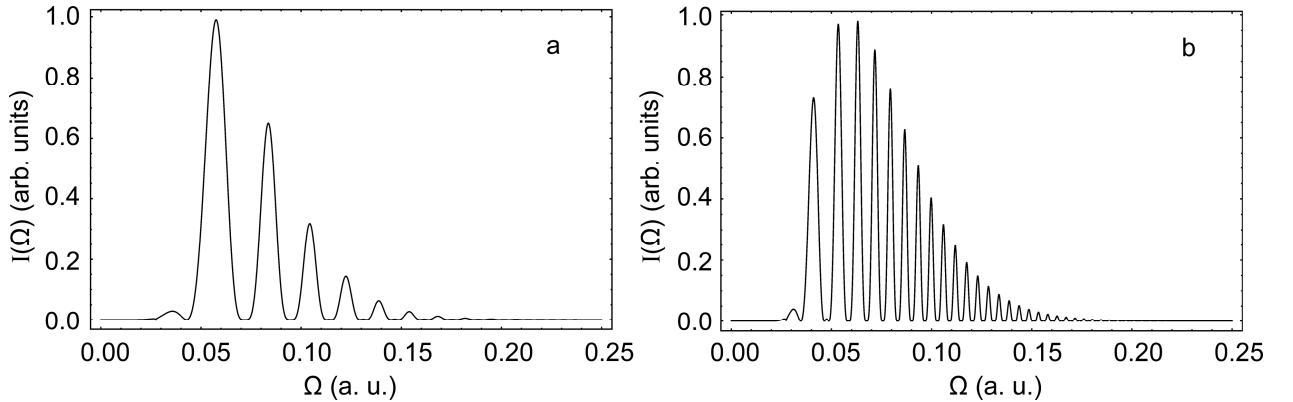

**Fig. 2.** The dependence of the recombination spectrum from the intensity of a laser field. The distance between centers of ionization and recombination is 5 a.u., the intensity of a laser field $I = 10^{11}$ W/cm$^2$ (a), $I = 2 \times 10^{10}$ W/cm$^2$ (b).

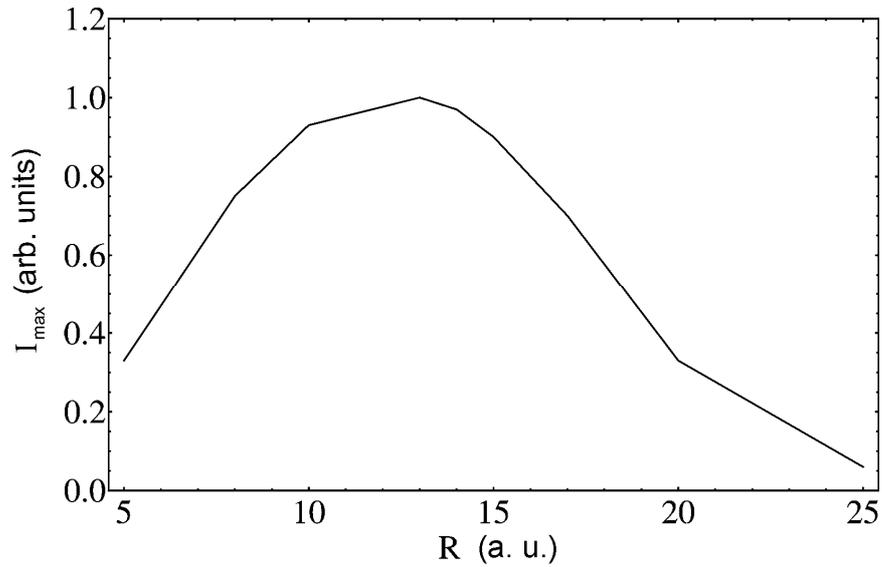

**Fig. 3.** The dependence of a maximum of the intensity of the photorecombination radiation from the distance between the centers of ionization and recombination for the intensity of the $CO_2$ laser $I = 10^{12}$ W/cm$^2$.

The figure 3 demonstrates the dependence of the photorecombination efficiency as a function of the distance between two attraction centers. The spectrum maximum lies at $\Omega = RF_0 - E_f$. The increasing $R$ leads to the shorter wavelength radiation.

*Conclusions.* – The measurements of the recombination radiation allow one to estimate ionization potential or the distance between two centers. Practically the same scheme can be applied to the problem with two quantum dots, taking them as initial and final recombination centers. Obviously, the intensity of photorecombination in three dimension case strongly depends on the angle between polarization and the axis of symmetry for a system of the two quantum dots.